\documentclass[a4paper,11pt]{article}

\usepackage{epsfig}
\newcommand{\beq}{\begin{equation}}
\newcommand{\eeq}{\end{equation}}
\ifx\pdfoutput\undefined
\usepackage[dvips,bookmarks]{hyperref}	
\else
\usepackage{hyperref}	
\fi
\hypersetup{colorlinks,bookmarksopen,bookmarksnumbered,citecolor=verdes,
linkcolor=blus,pdfstartview=FitH,urlcolor=rossos}
\def\hhref#1{\href{http://arxiv.org/abs/#1}{#1}} 

\usepackage{color}
\definecolor{rosso}{cmyk}{0,1,1,0.4}
\definecolor{rossos}{cmyk}{0,1,1,0.55}
\definecolor{rossoc}{cmyk}{0,1,1,0.2}
\definecolor{blu}{cmyk}{1,1,0,0.3}
\definecolor{blus}{cmyk}{1,1,0,0.6}
\definecolor{bluc}{cmyk}{1,1,0,0.1}
\definecolor{verde}{cmyk}{0.92,0,0.59,0.25}
\definecolor{verdec}{cmyk}{0.92,0,0.59,0.15}
\definecolor{verdes}{cmyk}{0.92,0,0.59,0.4}

\newcommand{\be}{\begin{equation}}
\newcommand{\ee}{\end{equation}}
\newcommand{\br}{\begin{eqnarray}}
\newcommand{\bea}{\begin{eqnarray}}
\newcommand{\eea}{\end{eqnarray}}
\newcommand{\er}{\end{eqnarray}}
\newcommand{\ba}{\begin{array}}
\newcommand{\ea}{\end{array}}
\newcommand{\bi}{\begin{itemize}}
\newcommand{\ei}{\end{itemize}}
\newcommand{\bn}{\begin{enumerate}}
\newcommand{\en}{\end{enumerate}}
\newcommand{\bc}{\begin{center}}
\newcommand{\ec}{\end{center}}

\newcommand{\eq}[1]{eq.~(\ref{#1})}

\newcommand{\rfn}[1]{(\ref{#1})}

\makeatletter
\def\art{\@ifnextchar[{\eart}{\oart}}
\def\eart[#1]#2#3#4#5#6{{\rm #2}, {#3 #4} {\rm (#6) #5} [{\hhref{#1}}]}
\def\hepart[#1]#2{{\rm #2, \hhref{#1}}}
\newcommand{\oart}[5]{{\rm #1}, {#2 #3} {\rm (#5) #4}}

\def\gappeq{\mathrel{\rlap {\raise.5ex\hbox{$>$}}
{\lower.5ex\hbox{$\sim$}}}}

\def\lappeq{\mathrel{\rlap{\raise.5ex\hbox{$<$}}
{\lower.5ex\hbox{$\sim$}}}}

\begin{document}
\begin{center}
{\LARGE {\color{rossos}\bf The golden ratio prediction\\ for the solar neutrino mixing}} \\
\vspace*{1cm}
{\bf  Yuji Kajiyama}$^a$,  {\bf Martti Raidal}$^a$ and  {\bf Alessandro Strumia}$^b$
\vspace{0.3cm}

{\em 
$^a$
National Institute of Chemical Physics and Biophysics, Ravala 10,
Tallinn 10143, Estonia \\
$^b$ Dipartimento di Fisica dell'Universit\`a di Pisa and INFN, Pisa, Italia}

\bigskip\bigskip

\centerline{\large\bf Abstract}
\begin{quote}\large

We present a simple texture that predicts the cotangent of the
solar neutrino mixing angle to be equal to the golden ratio.
This prediction is $1.4\sigma$ below the present best-fit value
and final SNO and KamLAND data could discriminate it from  tri-bi-maximal mixing.
The neutrino mass matrix is invariant under a Z$_2\otimes {\rm Z}'_2$ symmetry:
that geometrically is a reflection along the diagonal of the golden rectangle. 
Assuming an analogous structure in the quark sector suggests a golden
prediction for the Cabibbo angle, $\theta_C = \pi/4-\theta_{12}\approx 13.3^\circ$,
up to the uncertainties comparable to $V_{ub}$.

\end{quote}

\end{center}

\bigskip\bigskip


The neutrino mass matrix exhibits large mixing angles and a mild mass hierarchy.
If flavor has an underlying simplicity, it might be easier to recognize it from
neutrinos, rather than from charged leptons and quarks,
where it is difficult to disentangle ${\cal O}(1)$ factors from 
whatever generates their large mass hierarchies.
For the moment the atmospheric mixing angle $\theta_{23}$ is consistent with maximal
and the $\theta_{13}$ mixing angle is consistent with zero~\cite{nu}.
The solar neutrino mixing angle is measured to be
 $\tan^2\theta_{12}=0.445\pm0.045$~\cite{nu}:
large but not maximal.

Some  non-trivial information might therefore be contained in  $\theta_{12}$,
and several speculations have been invoked to explain its  value.
Tri-bi-maximal neutrino mixing~\cite{Harrison} is the most popular proposal.
It is based on the idea that there are both bi-maximal $(0,1,1)/\sqrt{2}$ 
as well as tri-maximal $(1,1,1)/\sqrt{3}$ mixings in the lepton sector.  
This proposal was initially not based on any deep argument,
except that it predicts a solar mixing angle $\tan^2 \theta_{12}=0.5$
close to the observed one, and more precisely 
1.2$\sigma$ above the experimental 
best fit value. 
Models that try to explain tri-bi-maximal mixing in terms of {broken} flavour symmetries
have been realized later~\cite{A4}.

We propose the following texture for the neutrino Majorana mass matrix $m_\nu$
and for the charged lepton Yukawa couplings $\lambda_E$:
\beq\label{m1}
m_\nu = \left(
\begin{array}{ c  c c}
0 &m&0 \\
m &m&0\\
0&0&m_{\rm atm}
\end{array}
\right), \;\;\;\;\;\;
\lambda_E = \pmatrix{\lambda_e & 0 &0 \cr
0& \lambda_\mu/\sqrt{2} & 
\lambda_\tau/\sqrt{2}\cr
0& -\lambda_\mu/\sqrt{2} & 
\lambda_\tau/\sqrt{2}}.
\eeq
It just assumes some texture zeroes
and some strict equalities among different entries.
The solar neutrino mixing comes from the neutrino sector,
and the atmospheric mixing from the charged lepton sector.
The mass eigenstates of the neutrino mass matrix are
$$m_\nu \cdot \pmatrix{1\cr \varphi\cr 0}=m\varphi  
\pmatrix{1\cr \varphi\cr 0},\qquad
m_\nu\cdot \pmatrix{-\varphi\cr 1\cr 0}=-\frac{m}{\varphi}  
\pmatrix{-\varphi\cr 1\cr 0},\qquad
m_\nu \cdot\pmatrix{0\cr0\cr 1} = m_{\rm atm}\pmatrix{0\cr0\cr 1} ,
$$
where $\varphi = (1+\sqrt{5})/2 = 
 1 + 1/\varphi\approx 1.62$ is known as the golden ratio~\cite{Euclid}.
Thanks to its peculiar mathematical properties this  constant appears in various 
natural phenomena,  possibly including solar neutrinos. 
Indeed the (1,2) entries of our neutrino mass
matrix are proportional to the generator of the Fibonacci recursion, $(x,y)\to (y,x+y)$, 
that describes  various growth phenomena with common limit $y/x =\varphi$.
The three neutrino mixing angles obtained from \eq{m1} are
$\theta_{\rm atm}=\pi/4$, $\theta_{13}=0$ and,
more importantly (see also~\cite{seealso}), 
\bea
\label{pred1}
\tan^2 \theta_{12}=1/\varphi^2=0.382,\qquad \hbox{i.e.}\qquad
\sin^2 2\theta_{12} = 4/5,
\eea
in terms of the parameter $\sin^2 2\theta_{12}$ directly measured by 
vacuum oscillation experiments, such as KamLAND.
This prediction for $\theta_{12}$ is 1.4$\sigma$ below the experimental  best fit value.
 The solar mass splitting, measured to be 
 $\Delta m^2_{\rm sun}=(8.0\pm0.3)~10^{-5}\,{\rm eV}^2$~\cite{nu}, in our case is given by
 $\Delta m^2_{\rm sun}=+ \sqrt{5} m^2$,
so the lightest neutrino mass is 
$|m_{1}|=m/\varphi=
(3.7\pm0.1)\,{\rm meV}$, and
$m_2 = (9.7\pm0.4)\,{\rm meV}$.
The parameter $|m_{ee}|$, probed by $0\nu2\beta$ decay experiments~\cite{nu}, vanishes.
A  positive measurement of $\theta_{13}$
might imply that our prediction for $\theta_{12}$
suffers an uncertainty up to  $\theta_{13}.$

Within the see-saw \cite{seesaw} context, 
the texture in  \eq{m1}  can be e.g.\ realized from the following  texture
for right-handed neutrino masses $M$ and neutrino Yukawa couplings $\lambda_\nu$:
\beq M = \label{M1}
\left(
\begin{array}{ c  c c}
-M &M&0 \\
M &0&0\\
0&0&M_{\rm atm}
\end{array}
\right), \;\;\;\;\;\;
\lambda_\nu = \pmatrix{\lambda & 0 &0 \cr
0& \lambda & 0\cr
0&0 & \lambda_{\rm atm} },
\eeq
together with $\lambda_E$ as in \eq{m1}.
The $-$ sign depends on our convention for the phases of  $N_1,N_2$.

\begin{figure}[t]
\centerline{\epsfxsize = 0.5 \textwidth \epsffile{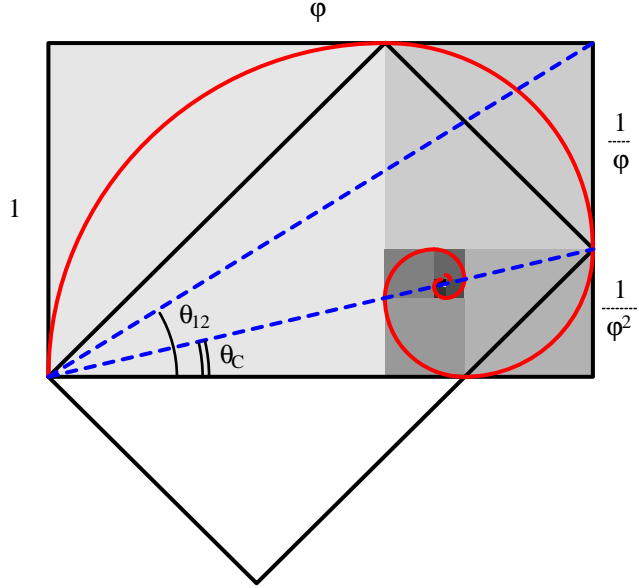}
}
\caption{\it Geometrical illustration of the connection between our predictions for
$\theta_{12} $ and $\theta_{C}$ and the golden rectangle.
The two dashed lines are the reflection axis of the ${\rm Z}_2$ symmetry
for the neutrino mass matrix and for the up-quark mass matrix.
\vspace*{0.cm}}
\label{f1}
\end{figure}

The simple form of the textures in \eq{m1}  and \eq{M1} suggests that the successful
prediction for $\theta_{12}$ may
be not just a numerical  coincidence, and motivates to realize it within flavor models,
that might give extra indications about the uncertain aspects discussed above.
The neutrino mass matrix $m_\nu$ is invariant under the
$L\to R \cdot L$ and  $L\to R' \cdot L$ symmetries, where
$L=(L_1,L_2,L_3)$ are the three left-handed leptons and
\beq \label{R}
R=\pmatrix{-1/\sqrt{5} & 2/\sqrt{5} & 0\cr
 2/\sqrt{5} & 1/\sqrt{5} & 0\cr
 0&0&1},
  \qquad
  R' = \pmatrix{1&0&0\cr 0&1&0\cr 0&0&-1},
  \label{sym}
  \eeq
are two Z$_2$ reflections, i.e.\
they satisfy $\det R= -1$, $R\cdot R^T=1$ and $R\cdot R=1$.
The first Z$_2$ is a reflection along the diagonal of the golden rectangle
in the $(1,2)$ plane, see Fig. \ref{f1}. The second Z$'_2$ is the $L_3 \to -L_3$ symmetry.
Any neutrino mass matrix is invariant under a $Z_2\otimes{\rm Z}'_2$ symmetry; 
in our case this symmetry, written in the texture basis, has the relatively simple form  in \eq{R}.


One can justify \eq{M1} by demanding that right-handed neutrinos $N=(N_1,N_2,N_3)$
transform in the same way as
left-handed doublets, $N\to R N$ and $N\to R' N$.
More precisely, these symmetries only force
 $m_\nu$ and $\lambda_\nu$, as well as the kinetic matrices for $L$ and $N$,
 all to have the same form as $M$ in \eq{M1}, up to
 an additional term proportional to the unit matrix.
This form of $M$ and $\lambda_E$ is more general than the simplest form in\eq{m1},
but is enough to imply the prediction for $\theta_{12}$,
 provided that $\lambda_E$ has the form in \eq{m1}.
Therefore the occurrence  of 
$0\nu2\beta$ decay is {\it allowed}  by the symmetries \rfn{sym} and
the neutrino mass eigenvalues are not restricted to satisfy $|m_2/m_1| = \varphi^2$.
This allows us to extend the analogous symmetries to the quark sector with very large 
mass hierarchies.


The Z$_2\otimes {\rm Z}'_2$ cannot be extended to right-handed leptons
$E=(E_1,E_2,E_3)$ in such a way that $\lambda_E$ has the form in \eq{m1}.
Indeed, there is a generic no-go argument that tells that the prediction
$\theta_{23}=\pi/4$ cannot be implied by any unbroken symmetry~\cite{Feruglio}:
one needs to consider suitably broken symmetries, and this opens a plethora of possibilities.
In our case, the lepton Yukawa matrix $\lambda_E$ has a simple form, diagonal up to a rotation by $\pi/4$ in the (2,3) sector.
Therefore such a simple structure of $\lambda_E$ could be explained in terms of an appropriate pattern of symmetry breaking.
Indeed one can extend the Z$_2\otimes {\rm Z}'_2$ to $E$ 
such that in the unbroken limit one has $\lambda_E=0$
(for example, by assuming  $E\to - E$ under both Z$_2$), 
and try to obtain the desired
small Yukawa couplings of charged leptons by suitably choosing symmetry-breaking 
effects, e.g.\ by demanding that the Z$_2\otimes{\rm Z}'_2$ symmetry is spontaneously broken
by vev of scalar fields that preferentially couple to well defined flavors,
$L_e$ and $L_\mu\pm L_\tau$.

If one imposes only the first Z$_2$ and disregards Z$'_2$, 
extra off-diagonal entries $m_{13} = m_{23}/\varphi$ are
allowed in the neutrino mass matrix. 
This does not affect $\theta_{12}$ and produces a non-zero $\theta_{13}$
correlated with a deviation of $\theta_{23}$ from maximal mixing:
neglecting the solar masses, and at first order in $\theta_{13}\ll 1$, the relation is
\beq \theta_{13}\simeq  (\theta_{23}-\pi/4)/\varphi.\eeq 
Better models can maybe be built using the icosahedral group,
isomorphic to A$_5$, the group of even permutations of 5 objects.
Indeed this group is connected to $\varphi$, as can be seen
either geometrically, or by noticing that the permutation factor $\omega=e^{2\pi i /5}$
satisfies $\varphi= 1 + \omega+\omega^4$.
The A$_4$ group has been used to justify bi-tri-maximal mixing,
because it allows to justify factors of $\omega = e^{2\pi i/3}$ in the mass matrices~\cite{A4}.


%

Noticing that the prediction \rfn{pred1} satisfies with high accuracy the 
`quark/lepton complementarity'~\cite{Smirnov,Raidal}, i.e.\  the observation that
$\theta_{12}+\theta_C $ is numerically close to $\pi/4$,
where $\theta_C$ is the Cabibbo angle, motivates us to give a golden geometric
explanation also to the Cabibbo angle.
As previously discussed, $\sin^2 2\theta_{12}=4/5$ arises if
$m_\nu$ is symmetric with respect to the
reflection around the diagonal of golden rectangle in the (12) plane,
and $\lambda_E$ is diagonal in the first two generations.
SU(5) unification relates the down-quark Yukawa matrix $\lambda_D$ to $\lambda_E$
and suggests that the up-quark Yukawa matrix $\lambda_U$ is symmetric, like $m_\nu$.
We therefore assume that $\lambda_D$ is diagonal in the two first generations and that
$\lambda_U$ is invariant under a Z$_2$ reflection described by a matrix analogous to
$R$ in \eq{R}, but with the factors $1\leftrightarrow 2$ exchanged.
Fig.~\ref{f1} illustrates the geometrical meaning of two reflection axis (dashed lines):
the up-quark reflection is along the diagonal of the golden rectangle tilted by $\pi/4$;
notice also the connection with the decomposition of the golden rectangle as
an infinite sum of squares (`golden spiral').
Similarly to the neutrino case, this symmetry allows for
two independent terms that can be tuned such that $m_u \ll m_c$:
\beq
\lambda_U= \lambda  \left(
\begin{array}{ c  c  c}
1 &0  &0\\
0 &1 &0 \\0 &0 &1 \\
\end{array}\right)  +
\frac{\lambda}{\sqrt{5}}
\left(
\begin{array}{ c  c  c}
-2 &1 &0\\
1 & 2 &0\\
0 & 0 & c\\
\end{array}
\right).
\eeq
The second term fixes $\cot\theta_C=\varphi^3$, as can be geometrically seen from fig.~\ref{f1}.
We therefore have
\beq \sin^2 2\theta_C = 1/5\qquad
\hbox{i.e.}\qquad \theta_{12}+\theta_C = \pi/4\qquad\hbox{i.e.}\qquad
V_{us}=\sin\theta_C = (1+\varphi^6)^{-1/2}= 0.229.\eeq
This prediction is $1.9\sigma$ above the present best-fit value,
$\sin \theta_C = 0.2258\pm 0.0021$~\cite{PDG}.
We expect that the golden prediction for $V_{us}$ has an uncertainty at least comparable to
$|V_{ub}|\sim |V_{td}|\sim\hbox{few}\cdot 10^{-3}$.

The textures we proposed are not stable under quantum corrections
(this is why flavor symmetries behind them must be broken), 
and presumably hold at some high scale of flavour generation, 
possibly of order $M_{\rm GUT}$ or $M_{\rm Pl}$.
Within the Standard Model, RGE corrections to $\theta_{12}$ and $\theta_{23}$ 
are numerically small,  at the per-mille level, because
suppressed by $\lambda_\tau^2$.  
The RGE corrections from new physics 
are model dependent and could be larger in models with larger
flavor dependent Yukawas~\cite{Antusch}, such as supersymmetric seesaw model 
at large $\tan\beta$.

 \bigskip
 
In conclusion, we have presented a simple texture (as well as its seesaw model realization) 
that might be behind the non-trivial value of the solar mixing angle.
The prediction $\theta_{12}=31.7^\circ$ is 1.4$\sigma$ below the 
present best-fit value and $2.6\sigma$ away from pure tri-bi-maximal mixing.
These two possibilities might be discriminated in the near future, as the
SNO (phase 3) and KamLAND experiments already collected new data that should
reduce the uncertainty on $\theta_{12}$~\cite{Petcov}\footnote{Including new KamLAND data presented
by I. Shimizu at the 10th Conference on Topics in Astroparticle and Underground Physics,
the global fit of KamLAND and SNO data now is $\tan^2\theta_{12}=0.49\pm0.06$
assuming $\theta_{13}=0$.}.
We identified a  Z$_2\otimes {\rm Z}'_2$  symmetry behind the neutrino mass matrix,
related to a reflection of the first two generations along the diagonal
of the golden rectangle.
This symmetry allows for an extra flavour-diagonal contribution to the mass matrices,
and consequently for $0\nu2\beta$ decay.
Assuming that a similar golden rectangle structure
also controls flavor mixing among the first two generation of quarks,
we get a geometric prediction for the Cabibbo angle,
$\theta_C = \pi/4-\theta_{12}=13.3^\circ$.
Although the golden predictions suffer from the uncertainties related to 
$(1,3)$ mixings in both sectors, we quoted those values with 3 digits because they
might optimistically hold up to few per mille precision.

\paragraph{Acknowledgements}
We thank Vincenzo Chilla,
Ferruccio Feruglio, Gino Isidori and Ernest Ma for communications. 
This work was supported in part by the ESF under grant No. 6140 and postdoc
contract 01-JD/06.

\end{document}